\documentclass[twocolumn,showpacs,floatfix,superscriptaddress,amsmath,amssymb,prl]{revtex4}
\usepackage{graphicx}
\usepackage{hyperref}
\usepackage{ulem}

\newcommand{\be}{\begin{equation}}
\newcommand{\ee}{\end{equation}}

\begin{document}

\title{Fidelity and quantum phase transitions}

\author{Huan-Qiang Zhou}
\affiliation{Department of Physics, Chongqing University, Chongqing
400044, The People's Republic of China} \affiliation{School of
Physical Sciences, University of Queensland, Brisbane, Qld 4072,
Australia}
\author{John Paul Barjaktarevi\v{c}}
\affiliation{School of Physical Sciences, University of Queensland,
Brisbane, Qld 4072, Australia}

\begin{abstract}
It is shown that the fidelity, a basic notion of quantum information
science, may be used to characterize quantum phase transitions,
regardless of what type of internal order is present in quantum
many-body states. If the fidelity of two given states vanishes, then
there are two cases: (1) they are in the same phase if the
distinguishability results from irrelevant local information; or (2)
they are in different phases if the distinguishability results from
relevant long-distance information. The different effects of
irrelevant and relevant information are quantified, which allows us
to identify unstable and stable fixed points (in the sense of
renormalization group theory). A physical implication of our results
is the occurrence of the orthogonality catastrophe near the
transition
points.
\end{abstract}

\date{\today}

\maketitle

In recent decades, significant advances have been achieved in the
study of quantum phase transitions (QPTs), both theoretically and
experimentally, in systems such as high-$T_c$ superconductors,
fractional quantum Hall liquids, and quantum magnets~\cite{sachdev}.
Conventionally, QPTs are characterized by singularities of the
ground state energy; first-order QPTs are characterized by
discontinuities in the first derivative of the energy, whereas
second-order (higher-order) QPTs are characterized by
discontinuities in the second (higher) derivative of the energy. At
singular points, the spectrum is gapless.

The focus of the traditional description of QPTs in condensed matter
physics is a Hamiltonian and its spectrum. The most studied QPTs fit
into the conventional Landau-Ginzburg-Wilson paradigm. A central
concept is a local order parameter, whose non-zero value
characterizes a symmetry-breaking phase, a unique feature which only
exists for a system with infinite number of degrees of freedom, in
contrast to QPTs resulting from a level crossing, which may happen
in a finite size system.
However, there
exist phases that are not described by symmetry-breaking orders,
which results in continuous QPTs beyond the Landau-Ginzburg-Wilson
paradigm~\cite{wen}. Indeed, such phase transitions exists between
two phases with the \textit{same} symmetry~\cite{wu}, between two
states with \textit{incompatible} symmetries~\cite{senthil}, and
even between two states with symmetry breaking, if the simultaneous
changes of topological/quantum orders occur~\cite{ran}.

On the other hand, quantum information science brings about an
emerging picture which studies QPTs from the ground state wave
functions of systems. The cross fertilization of quantum many-body
theory and quantum information science has led to fruitful outcomes.
One aspect is the study of the possible role of entanglement in
characterizing QPTs~\cite{preskill,osborne,vidal,levin}. Remarkably,
for quantum spin chains, the von Neumann entropy, as a bipartite
entanglement measure, exhibits qualitatively different behaviors at
and off criticality~\cite{vidal}. Further, QPTs in spin chains
characterized by local Hamiltonians with matrix product ground
states exhibit a different type of QPTs from the standard paradigm,
where the ground state energy remains analytic~\cite{wolf}.

In this Letter, we investigate the role of the fidelity, a basic
notion of quantum information science, in characterizing QPTs. As a
distance measure, the fidelity describes how close two given quantum
states are. Therefore, it is natural to expect that the fidelity may
be used to characterize drastic changes in quantum states when
systems undergo QPTs, regardless of what type of internal order is
present in quantum many-body states. It is shown that the fidelity
of two states vanishes due to different reasons depending on whether
they are in the same phase or in different phases. For the former,
the orthogonality results from irrelevant local information; for the
latter, the orthogonality results from relevant long-distance
information.
One may identify unstable and stable fixed points by quantifying the
effects of irrelevant and relevant information. We present examples
exhibiting a second order phase transition, a critical line, a level
crossing and an infinite order phase transition. An implication of
our results is the orthogonality catastrophe near the transition
points~\cite{anderson}. It is proper to stress here that Zanardi and
Paunkovi\'{c}~\cite{zanardi} were the first to exploit the ground
state overlap, which is equivalent to the fidelity at zero
temperature, to detect QPTs in the Dicke model and the $XY$ spin
chain.

We consider a quantum system $S$
described by a Hamiltonian $H(\lambda)$, with $\lambda$ a control
parameter~\cite{multiple}. The latter may be tuned to drive the
system to undergo a QPT at some transition point $\lambda_c$. It
should be stressed that the transition point may be of any type,
caused by a level crossing, spontaneous symmetry breaking, or the
occurrence of some exotic orders such as quantum/topological orders.
Here we stress that the size of the system may be either
thermodynamically large or finite, depending on the type of QPT. We
first consider a system at zero temperature. Then the system is in a
ground state. The main conclusion is as follows:

{\textit{Proposition 1.}} If a quantum system $S$ described by the
Hamiltonian $H(\lambda)$ undergoes a QPT at a transition point
$\lambda_c$, with $|\psi\rangle$ and $|\phi\rangle$ denoting,
respectively, representative ground states in the phases $\lambda >
\lambda_c$ and $\lambda < \lambda_c$, then we have
$\langle\psi|\phi\rangle=0$. That is, the fidelity $F(\psi, \phi)
\equiv | \langle \psi | \phi \rangle |$ of two representative states
from different phases vanishes. Conversely, if the fidelity of two
given states $|\psi\rangle$ and $|\phi\rangle$ vanishes, i.e., they
are orthogonal to each other,
then either (1) they are in two different phases if the
orthogonality results from relevant long-distance information
present in the states; or (2) $|\psi\rangle$ and $|\phi\rangle$
belong to the same phase if the orthogonality results from
irrelevant local (short-distance) information.

{\textit{Proof.}}  A representative state $|\psi\rangle$ in a given
phase is reliably distinguishable from a representative state
$|\phi\rangle$ in the other phase, due to the occurrence of
different orders in different phases~\cite{levelcross}. Here the
distinguishability is understood in the sense of quantum
measurements~\cite{nielsen}. On the other hand, two non-orthogonal
states cannot be reliably distinguished~\cite{nielsen}, as follows
from the basic Postulate of Quantum Mechanics on quantum
measurements. This implies that $\langle\psi|\phi\rangle=0$.
Conversely, if $|\psi\rangle$ and $|\phi\rangle$ are orthogonal to
each other, then they are reliably distinguishable. Therefore, there
must be some physical observable which may be exploited to
distinguish the states and one may take this observable as an order
parameter~\cite{oshikawa}. Therefore whether or not the two states
are in the same phase depends on whether the difference unveiled by
such an order parameter is quantitative or qualitative.  In the
former case, the distinguishability results from the short-distance
details of the system, i.e., irrelevant local information of the
system. Since the long-distance behavior of a system does not depend
on the short-distance details of the system~\cite{wen}, so the two
states share the same long-distance information and must be in the
same phase. In the latter case, the distinguishability results from
relevant long-distance information of the system, so the two states
must belong to different phases.

To complete our characterization of QPTs in terms of the fidelity,
it is necessary to quantify the different effects of irrelevant and
relevant information. To this end, we restrict ourselves to consider
systems exhibiting QPTs with symmetry-breaking orders and study the
scaling behaviors of the fidelity with the system size $L$. We have

{\textit{Proposition 2.}} For a quantum system of a large size $L$,
the fidelity $F(\lambda, \lambda')$ scales as $[d(\lambda,
\lambda')]^L$~\cite{definition}, with $0 \leq d \leq 1$ some
constant characterizing how fast the fidelity changes when the
thermodynamic limit is approached. If $\lambda$ and $\lambda'$ are
in different phases, then $d(\lambda_<, \lambda_>) \geq d(\lambda,
\lambda')\geq d(\lambda_1, \lambda_2)$, where $\lambda_1$ and
$\lambda_2$ are fixed points to which quantum states in two
different phases flow under renormalization group (RG)
transformations, and  $\lambda_<$ and $\lambda_>$ approach the
transition value $\lambda_c$ from both sides (in the thermodynamic
limit). If $\lambda$ and $\lambda'$ are in the same phase, then
$d(\lambda, \lambda')\rightarrow 1$ as $\lambda\rightarrow\lambda'$.

{\textit{Proof.}} It is well-known that symmetry-breaking orders
emerge only in the thermodynamic limit. Therefore, the fidelity for
any two ground states of a finite size system does not vanish, but
is exponentially small due to the locality of order parameters when
the size $L$ is very large. Suppose the system flows to two
different stable fixed points $\lambda_1$ and $\lambda_2$ under RG
transformations~\cite{cardy}. Then we have $d(\lambda, \lambda')
\geq d(\lambda_1,\lambda_2)$ if $\lambda$ and $\lambda'$ are in
different phases, and $\lambda$ and $\lambda'$ flow to $\lambda_1$
and $\lambda_2$, respectively. This follows from the fact that two
states at stable fixed points $\lambda_1$ and $\lambda_2$ are the
most distinguishable states, since there is no suppression in order
parameters caused from quantum fluctuations.  The irrelevant
short-distance information present in the states $\psi(\lambda)$ and
$\psi(\lambda')$ only makes them less distinguishable. Indeed, two
states in different phases possess different relevant long distance
information, so they become more distinguishable when the
short-distance information is washed away. The maximum is reached
when $\lambda$ and $\lambda'$ approach the transition point
$\lambda_c$ from both sides, since strong quantum fluctuations makes
them relatively less distinguishable among states in different
phases. Thus $d(\lambda, \lambda') \leq d(\lambda_<, \lambda_>)$.
However, $d(\lambda_<, \lambda_>)$ must be less than 1, because the
short-distance (irrelevant) information cannot undo what results
from the long-distance (relevant) information. In contrast,
$d(\lambda, \lambda')$ smoothly approaches $1$ if $\lambda$
approaches $\lambda'$ in the same phase. In fact, states in the same
phase enjoy exactly the same relevant long-distance information, so
they are indistinguishable at long-distance scales. However, the
short-distance information makes them distinguishable. More
precisely, $d(\lambda,\lambda')$ decreases with $\lambda$ along a
renormalization group flow for a fixed $\lambda'$, if $\lambda$ and
$\lambda'$ are in different phases, and $d(\lambda,\lambda')$
increases with $\lambda$, until it reaches $1$ when $\lambda
=\lambda'$, then it decreases with $\lambda$ along a renormalization
group flow for a fixed $\lambda'$, if $\lambda$ and $\lambda'$ are
in the same phase. Since $d(\lambda,\lambda')$ is continuous, and it
is symmetric under exchange $\lambda \leftrightarrow \lambda'$, one
may recognize unstable and stable fixed points as pinch
points~\cite{pinch} and global minima, respectively.

\textit{Remark 1.} One may extend the above argument to other types
of QPTs. However, the exponential scaling of the fidelity with the
size $L$ is not necessarily valid. Actually, for QPTs resulting from
level crossings, $d$ only takes values 0 or 1. For QPTs in matrix
product systems~\cite{wolf}, the fidelity may exhibit a fast oscillating
scaling behavior with an exponentially decaying envelope. Therefore
\textit{different scaling behaviors signal different types of QPTs}.

\begin{figure}[ht]
\vspace*{4.0cm} \includegraphics{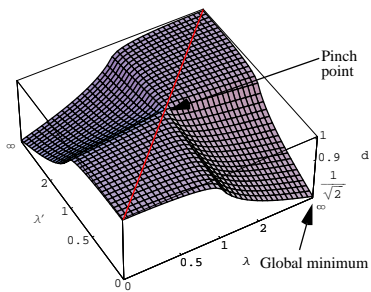}
  \caption{(Color online)
  The scaling parameter $d$, which appears in the fidelity
  scaling $F(\lambda,\lambda') \sim d^L$, for the two
states of quantum transverse Ising model as a function of $\lambda$
and $\lambda'$. The transition point $\lambda_c=1$ is characterized
as a pinch point ($1,1$) and the two stable fixed points, to which
all states in two phases flow, are characterized as the global
minima at ($0,\infty$) and ($\infty,0$). The red line denotes $
d(\lambda,\lambda)=1$.}
  \label{fidelity}
\end{figure}

Now we turn to the finite temperature case. Then the system is in a
mixed state, characterized by a density matrix $\rho$. A mixed state
may be purified~\cite{nielsen} if one introduces a reference system
$R$, which has the same state space, e.g., another copy of $S$. That
is, define a pure state $|SR\rangle$ for the joint system $SR$ such
that $\rho = \textrm{tr}_R (|SR\rangle\langle SR|)$. Assume that
orders present in ground states survive thermal fluctuations. Then
we have

{\textit{Proposition 3.}} If $\rho$ and $\sigma$ denote,
respectively, two representative states of different phases for a
quantum system $S$ at finite temperature, then the fidelity
$F(\rho,\sigma) = \textrm{tr} \sqrt {\rho^{1/2} \sigma \rho^{1/2}}$
vanishes. Conversely, if the fidelity of two states $\rho$ and
$\sigma$ is zero, i.e., $F(\rho,\sigma)=0$, then either (1) $\rho$
and $\sigma$ are in two different phases and the orthogonality
results from the qualitative difference of the relevant
long-distance information; or (2) $\rho$ and $\sigma$ belong to the
same phase and the orthogonality results from the quantitative
difference of the irrelevant local information.

{\textit{Proof.}}  There is an alternative characterization of the
fidelity due to Uhlmann's theorem~\cite{uhlmann}, which states that
$F(\rho,\sigma) = \textrm{max}_{|\psi\rangle,|\phi\rangle}
|\langle\psi|\phi\rangle|$, where the maximization is over all
purifications $|\psi\rangle$ of $\rho$ and $|\phi\rangle$ of
$\sigma$ in a joint system $SR$, with $R$ being a copy of $S$. Since
\textit{any} purifications of $\rho$ and $\sigma$ must be orthogonal
to each other, due to the fact that different orders in different
phases make them reliably distinguishable. Combined with Uhlmann's
theorem, one concludes that $F(\rho,\sigma)=0$. Conversely, if
$F(\rho,\sigma)=0$, Uhlmann's theorem implies that the purifications
of $\rho$ and $\sigma$ must be orthogonal to each other. Thus they
must be reliably distinguishable. Since the reference system $R$ is
fictitious, and only the system $S$ itself is accessible, so there
must be some physical observable in the system $S$ to characterize
the distinguishability of the purifications. Thus we are led to
conclusions along the same line of reasoning as {\textit
{Proposition 1}}.

\textit{Remark 2.} The angle $A$ between states $\rho$ and $\sigma$
defined by $A(\rho,\sigma) \equiv \arccos F(\rho,\sigma)$ is a
metric~\cite{nielsen}. That is, it is non-negative, symmetric and is
equal to zero if and only if $\rho =\sigma$, and it obeys the
triangle inequality: $A(\rho,\tau) \leq A(\rho,\sigma) + A(\sigma,
\tau)$ for any states $\rho, \sigma$ and $\tau$. For brevity, we
consider a finite (but large) size system at zero temperature. The
angle $A$ between two states at $\lambda$ and $\lambda'$, if one
partitions the interval $[\lambda, \lambda']$ into $N$ small pieces,
satisfies $A\leq N A_N$, with $A_N$ being the largest angle for $N$
different pieces. If the two states are in different phases, then
there exists at least one piece, for which the angle for the states
at the ends of the piece is greater than $A/N$. Indeed, the angle
for the piece across the transition point is $\sim \pi/2$. Instead,
if the two states are in the same phase, then the interval may be
partitioned into $N$ pieces such that the angle between the ends of
each piece is $A/N$, which can be as small as required.

One of physical implications of the above results is the Anderson
orthogonality catastrophe around the transition point $\lambda_c$
~\cite{anderson}. Indeed, no matter how close $\lambda_<$ and
$\lambda_>$ are to the transition value $\lambda_c$, the
corresponding ground states $|\psi(\lambda_<)\rangle$ and
$|\psi(\lambda_>)\rangle$ must be \textit{orthogonal} to each other,
i.e., $\langle\psi(\lambda_<)|\psi(\lambda_>)\rangle=0$ (in the
thermodynamic limit). This was first discovered by Zanardi and
Paunkovi\'{c}~\cite{zanardi} for the Dicke model and the $XY$ spin
chain, who also pointed out that the orthogonality catastrophe may
reveal itself in the Loschmidt echo~\cite{loschmidt}.

{\it Quantum $XY$ spin chain.} The quantum $XY$ spin chain is
described by the Hamiltonian
\begin{equation}
H= -\sum_{j=-M}^M ( \frac {1+\gamma}{2} \sigma^x_j \sigma^x_{j+1} +
\frac {1-\gamma}{2} \sigma^y_j \sigma^y_{j+1}  + \lambda \sigma^z_j
). \label{HXY}
\end{equation}
Here $\sigma_j^x, \sigma_j^y$, and $\sigma_j^z$  are the Pauli
matrices at the $j$-th lattice site. The parameter $\gamma$ denotes
an anisotropy in the nearest-neighbor spin-spin interaction, whereas
$\lambda$ is an external magnetic field. The Hamiltonian (\ref{HXY})
may be diagonalized as $H=\sum_k \Lambda_k (c_k^\dagger c_k -1)$,
where
$\Lambda_k=\sqrt {(\lambda-\cos(2\pi k/L))^2+\gamma^2 \sin^2 (2\pi
k/L)}$, with $c_k$ and $c_k^\dagger$ denoting free fermionic modes
and $L=2M+1$. The ground state $|\psi\rangle$ is the vacuum of all
fermionic modes defined by $c_k|\psi\rangle=0$, and may be written
as $|\psi\rangle=\prod^{M}_{k=1} (\cos (\theta_k/2)-i\sin
(\theta_k/2)c^\dagger_k c^\dagger_{-k})|0\rangle_k|0\rangle_{-k}$,
where $|0\rangle_k$ is the vacuum of the $k$-th mode, and $\theta_k$
is defined by $\cos \theta_k = (\cos (2\pi k/L)-\lambda)/\Lambda_k$.
Therefore, the fidelity $F$ for two states $|\psi (\lambda,
\gamma)\rangle$ and $|\psi (\lambda',\gamma')\rangle$ takes the
form,
\begin{equation}
F=\prod^M_{k=1}\cos \frac{\theta_k-\theta'_k}{2},
\end{equation}
where the prime denotes that the corresponding variables take their
values at $\lambda'$ and $\gamma'$. Obviously, $F=1$ if
$\lambda=\lambda'$ and $\gamma = \gamma'$. Generically, $\cos
\frac{\theta_k-\theta'_k}{2}  < 1$, therefore the fidelity decays
very fast when $\lambda$ and/or $\gamma$ separate, respectively,
from $\lambda'$ and/or $\gamma'$.

Let us first consider the Heisenberg $XX$ model in an external
magnetic field ($\gamma =0$), with a critical line characterized by
$\lambda \in [0,1)$. In this case, $\cos \theta_k=1$ if $\cos (2\pi
k/L) \geqslant \lambda$ and $\cos \theta_k=-1$ if $\cos (2\pi k/L) <
\lambda$. Therefore, if both $\lambda$ and $\lambda'$ is greater
than 1, then we have $F=1$, consistent with the fact that the
transition point $\lambda =1$ for the Heisenberg $XX$ model results
from a level crossing. If $\lambda
>1$ and $\lambda' \leqslant1$ or vice versa, then $F=0$, consistent
with \textit{Proposition 1}. Suppose $\lambda<1$ and $\lambda'<1$,
$F=1$ only if $\lambda$ and $\lambda'$ are so close that there is no
$k$ satisfying $\lambda <\cos (2\pi k/L) < \lambda'$ or $\lambda'
<\cos (2\pi k/L) < \lambda$. In the thermodynamic limit,  such a $k$
always exists irrespective of $\lambda$ and $\lambda'$. That is,
$F=0$ except for $\lambda=\lambda'$, indicating that there is a line
of critical points $[0,1)$, identified as the Luttinger liquids with
dynamical critical exponent $z=1$. We stress that the transition
point $\lambda_c=1$ with $z=2$ controls the global features of the
system.

Next consider the quantum transverse Ising universality class with
the critical line $\gamma\neq 0$ and $\lambda=1$. There is only one
(second-order) critical point $\lambda_c=1$ separating two gapful
phases: (spin reversal) $Z_2$ symmetry-breaking and symmetric
phases.  In the thermodynamic limit, the scaling parameter $d$ takes
the form: $\ln d(\lambda,\lambda') = 1/(2\pi) \int ^\pi_0 d\alpha
\ln {\cal F} (\lambda,\lambda';\alpha)$, where ${\cal F}
(\lambda,\lambda';\alpha) = \cos [\vartheta
(\lambda;\alpha)-\vartheta(\lambda';\alpha)]/2$, with $\cos
\vartheta (\lambda;\alpha) = (\cos \alpha - \lambda)/\sqrt {(\cos
\alpha -\lambda)^2+\gamma^2 \sin^2 \alpha}$. We plot $d$ in Figure 1
for the transverse Ising model ($\gamma=1$). One observes that the
transition point $\lambda_c=1$ is characterized as a pinch point
($1,1$) and that the two stable fixed points at $\lambda =0$ and
$\lambda =\infty$ are characterized as the global minima, which take
value $1/\sqrt 2$ at ($0,\infty$) and ($\infty,0$).

The last case is the disorder-line, i.e., a unit circle given by
$\lambda^2 + \gamma^2 =1$ in the $\lambda-\gamma$ plane. We plot the
scaling parameter $d$ in Figure 2, from which one may read off that
there are two transition points $(\pm 1, 0)$ and that there are two
phases corresponding to the upper and lower semi-circles, with $(0,
\pm 1)$ as stable fixed points. The latter corresponds to two states
with all spins aligning in the $x$ and $y$ directions, respectively.
The system is dual to a spin $1/2$ model with three-body
interactions
\begin{equation}
H= \sum_i 2(g^2 -1) \sigma^z_i \sigma^z_{i+1} - (1+g)^2 \sigma^x_i +
(g-1)^2 \sigma^z_i \sigma^x_{i+1} \sigma^z_{i+2},
\end{equation}
with $\lambda = (1-g^2)/(1+g^2), \gamma = 2g/(1+g^2)$. As shown in
Ref.~\cite{wolf}, the model exhibits a peculiar QPT in the
thermodynamic limit, with divergent correlation length, vanishing
energy gap, but analytic ground state energy. We emphasize that the
parameter space should be compactified by identifying $g=+\infty$
and $g=-\infty$, due to the fact that $H(+ \infty) =H(-\infty)$.
Since ground states are matrix product states~\cite{wolf}, it is
straightforward to get the fidelity $F$ for two states $|\psi
(g)\rangle$ and $|\psi (g')\rangle$,
\begin{equation}
F = \frac {|(1+\sqrt {g g'})^L + (1-\sqrt {g g'})^L|}{\sqrt
{[(1+g)^L + (1-g)^L][(1+g')^L + (1-g')^L]}}.
\end{equation}
The fidelity $F$ decays exponentially for two states in the same
phase, but it is oscillating very fast with exponentially decaying
envelope for two states in different phases. From this one may
extract the scaling parameter $d$ as $d(g,g') = \sqrt {1+|g g'|}/
\sqrt {(1+|g|)(1+|g'|)}$ if $g$ and $g'$ are in different phases,
and  $d(g,g') = (1+\sqrt {|g g'|})/ \sqrt {(1+|g|)(1+|g'|)}$ if $g$
and $g'$ are in the same phase. There are two transition points,
i.e., $g=0$ and $\infty$. All states for positive $g$ flow to the
product state ($g=1$) with all spins aligning in the $x$ direction,
and all states for negative $g$ flow to the cluster
state~\cite{raussendorf} ($g=-1$).

\begin{figure}[t]
\vspace*{4.0cm} \includegraphics{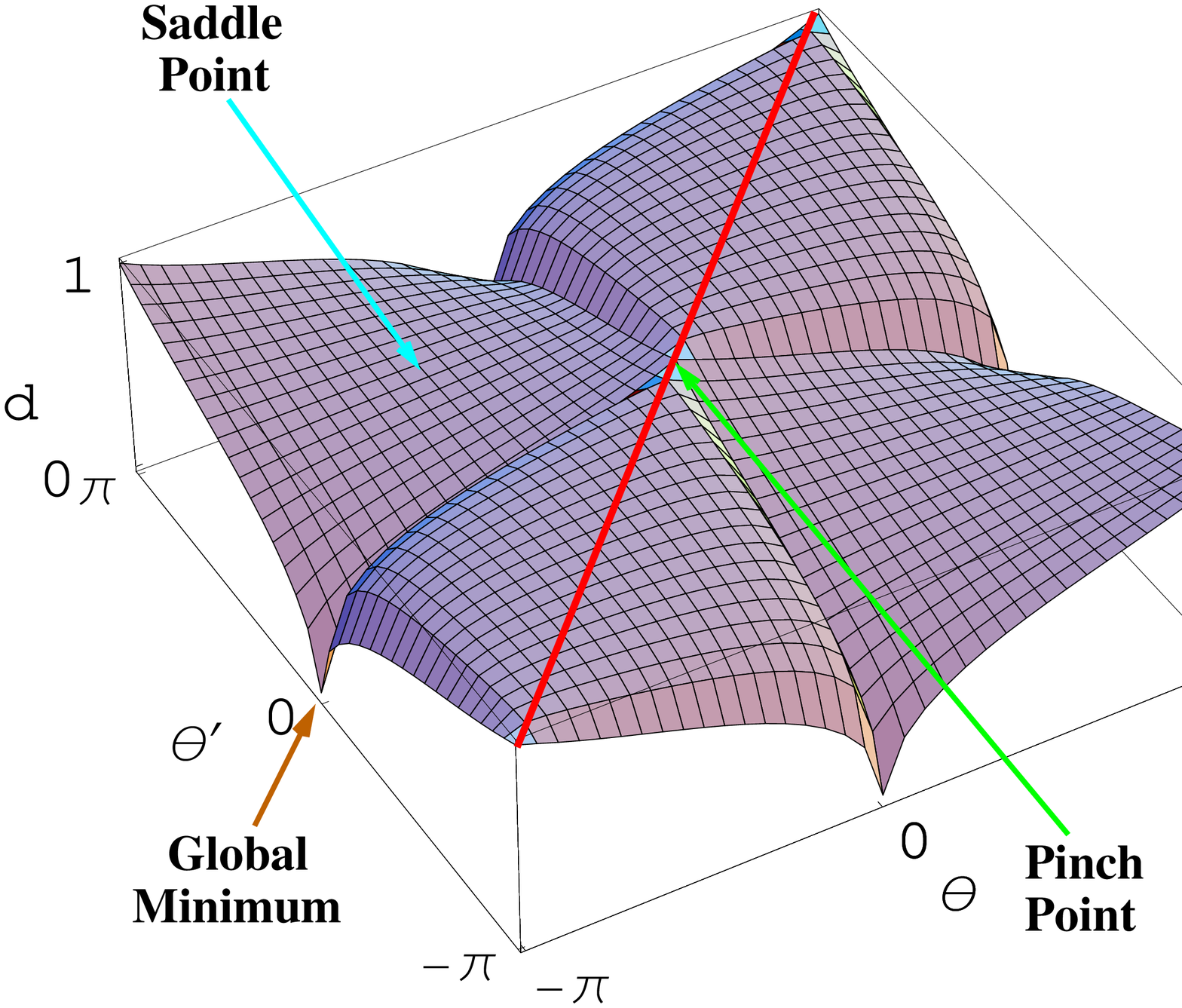}
\includegraphics{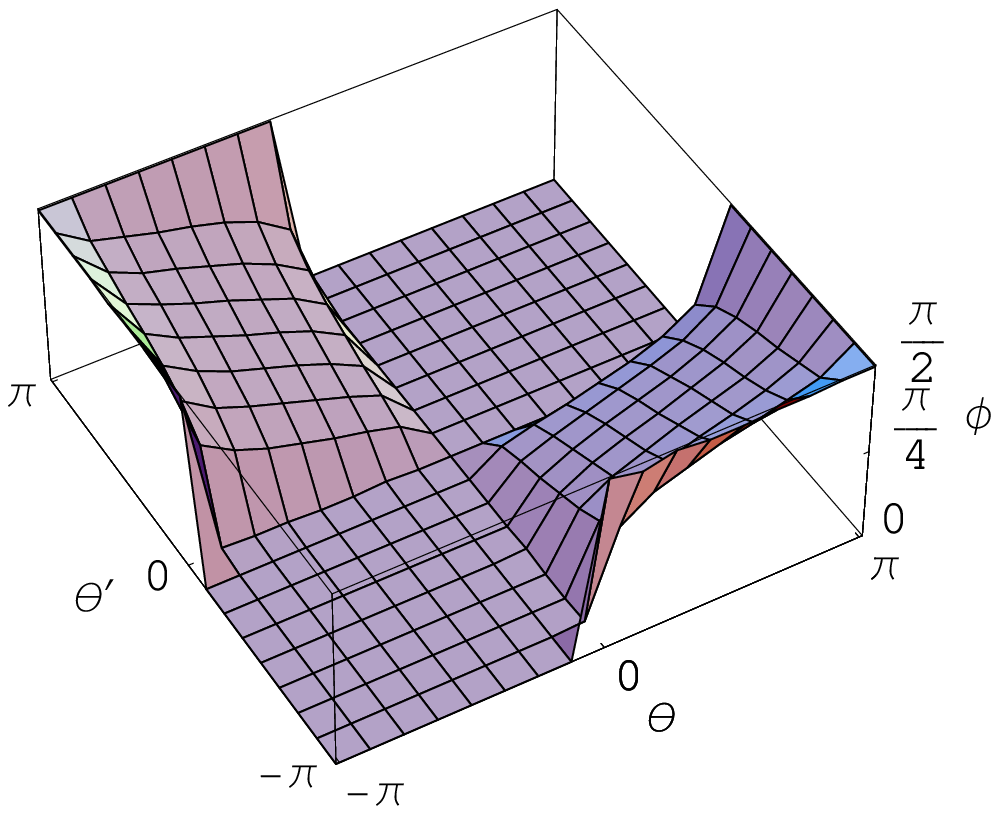}
  \caption{(Color online)
  For the disorder line, $\lambda^2 + \gamma^2 =1$ parameterized
as $\lambda=\cos\theta$, and $\gamma=\sin\theta$, the fidelity
$F(\theta,\theta')$ scales as $d^L cos (L\phi)$.  Thus for
$\phi\ne0$, there is a sinusoidal oscillation with an exponential
envelope in the fidelity.  Left: The scaling parameter $d$ as a
function of $\theta$ and $\theta'$, displaying pinch points at
$(\theta,\theta')=(0,0)$ and $(\pi,\pi)$, and saddle points at
$(\pi/2,-\pi/2)$ and $(-\pi/2, \pi/2)$, if one identifies $\pi$ with
$-\pi$. The pinch points characterize the two transition points,
while the saddle points characterize the stable fixed points, to
which all states in two phases flow. The global minima at $(0,\pi)$
and $(\pi, 0)$ correspond to the transition points, due to the fact
that both irrelevant and relevant information are different. Right:
The phase $\phi$ as a function of $\theta$ and $\theta'$. }
  \label{temp}
\end{figure}

In summary, we have established an intriguing connection between the
fidelity and QPTs in particular and  between quantum information
science and condensed matter physics in general.

We thank Paolo Zanardi, John Fjaerestad and Sam Young Cho for
helpful discussions and comments. This work is supported by
Australian Research Council.


\begin{thebibliography}{99}
\bibitem{multiple} The extension of our discussion to systems
depending on more than one parameter is straightforward.
\bibitem{levelcross} An exceptional case arises for a system with
a level-crossing, in which no order emerges.
However, in such a case, the states from
different phases are obviously orthogonal.
\bibitem{definition} Formally, $d(\lambda,\lambda')$ may be
defined as $\ln d(\lambda,\lambda') = \lim_{L \rightarrow \infty}
\ln F(\lambda,\lambda') /L$ in the thermodynamic limit.
\bibitem{pinch} We define a pinch point as an intersection of two singular
lines, along which four smooth pieces patch together.
\end{thebibliography}
\end{document}